\documentclass[useAMS,usenatbib]{mn2e}

\usepackage{epsfig}
\usepackage{amsmath}
\usepackage{amssymb}
\usepackage{mathrsfs}
\usepackage{graphicx}
\usepackage{relsize}
\usepackage{mathtools}
\usepackage{multirow}

\newcommand{\apj}{ApJ} 
\newcommand{\apjl}{ApJ} 
\newcommand{\aap}{A\&A} 
\newcommand{\araa}{ARAA} 
\newcommand{\mnras}{MNRAS} 

\title[The density variance -- Mach number relation]{The density variance -- Mach number relation in isothermal and non-isothermal adiabatic turbulence}
\author[C.~A.~Nolan, C.~Federrath and R.~S.~Sutherland]{
C.~A.~Nolan$^{1}$\thanks{E-mail: chris.nolan@anu.edu.au (ANU)},
C.~Federrath$^{1}$\thanks{E-mail: christoph.federrath@anu.edu.au (ANU)} \&
R.~S.~Sutherland$^{1}$\thanks{E-mail: ralph.sutherland@anu.edu.au (ANU)} \\
$^{1}$Research School of Astronomy $\&$ Astrophysics, The Australian National University, Canberra, ACT 2611, Australia}

\begin{document}

\maketitle

\begin{abstract}
The density variance -- Mach number relation of the turbulent interstellar medium is relevant for theoretical models of the star formation rate, efficiency, and the initial mass function of stars. Here we use high-resolution hydrodynamical simulations with grid resolutions of up to $1024^3$ cells to model compressible turbulence in a regime similar to the observed interstellar medium. We use Fyris Alpha, a shock-capturing code employing a high-order Godunov scheme to track large density variations induced by shocks. We investigate the robustness of the standard relation between the logarithmic density variance ($\sigma_s^2$) and the sonic Mach number ($\mathcal{M}$) of isothermal interstellar turbulence, in the non-isothermal regime. Specifically, we  test ideal gases with diatomic molecular ($\gamma = 7/5$) and monatomic ($\gamma = 5/3$) adiabatic indices. A periodic cube of gas is stirred with purely solenoidal forcing at low wavenumbers, leading to a fully-developed turbulent medium. We find that as the gas heats in adiabatic compressions, it evolves along the relationship in the density variance -- Mach number plane, but deviates significantly from the standard expression for isothermal gases. Our main result is a new density variance -- Mach number relation that takes the adiabatic index into account: $\sigma_s^2=\ln\left(1+b^2 \mathcal{M}^{(5\gamma+1)/3}\right)$ and provides good fits for $b\mathcal{M}\lesssim1$. A theoretical model based on the Rankine-Hugoniot shock jump conditions is derived, $\sigma_s^2 = \ln\{ 1 + (\gamma+1)b^2\mathcal{M}^2/[(\gamma-1)b^2\mathcal{M}^2+2]\}$, and provides good fits also for $b\mathcal{M}>1$. We conclude that this new relation for adiabatic turbulence may introduce important corrections to the standard relation, if the gas is not isothermal ($\gamma\ne1$).
\end{abstract}


\section{Introduction}

The interstellar medium (ISM) is a complex, turbulent, multi-phase gaseous medium, which permeates the space between stars in the galactic plane \citep{Ferriere:2001vr}. It is an essential part of the evolutionary cycle in stars, recycling the products of nucleosynthesis from dying stars and creating the stellar nurseries for a new generation of star formation \citep{MacLowKlessen2004,ElmegreenScalo2004,McKeeOstriker2007,Krumholz2014,PadoanEtAl2014}. The ISM interacts with supernova explosions, protostellar jets, winds and outflows, which shape its structure and drive the turbulence we observe via atomic and molecular line observations of the ISM.

In many simulations that include an ISM \citep[e.g.,][]{BlandHawthorn:2007wn,Wagner:2011io,Cooper:2009br,Fischera:2005uo}, a statistical construction with isotropic properties related to turbulent statistics have been used, as a proxy for the turbulent ISM. Causal models of the turbulent ISM will drastically increase the accuracy of these models, but this first involves an in-depth study of the statistics and evolution of fully-developed turbulence.

In purely isothermal gas, the probability density function (PDF) of the gas densities may be approximated by a lognormal distribution \citep{Vazquez1994}, which in log-space has the form
\begin{equation} \label{pdf}
p_\mathrm{LN}(s) = \frac{1}{\sqrt{2 \pi \sigma_{s}^2}} \exp \left[ - \frac{1}{2}  \frac{(s - \bar{s})^2}{\sigma_{s}^2} \right],
\end{equation}
where $s = \ln(\rho/\bar{\rho})$, $\bar{s}$ and $\sigma_{s}^2$ are the mean and variance of the logarithm of the density $\rho$, scaled by the mean density, $\bar{\rho}$. The logarithmic density variance is a function of the root-mean-squared (rms) sonic Mach number ($\mathcal{M}$), and is given by
\begin{equation} \label{relation1}
\sigma_s^2 = \ln\left(1+b^2 \mathcal{M}^2\right).
\end{equation}
The coefficient $b$ is known as the turbulence driving parameter and depends on the mode mixture induced by the turbulent forcing mechanism \citep{Federrath:2008ey}. Purely solenoidal (divergence-free) driving leads to $b=1/3$, while purely compressive (curl-free) driving corresponds to $b=1$.

Equation~(\ref{relation1}) has been studied extensively for isothermal gases \citep{Padoan:1997tn,Passot:1998wz,Kritsuk:2007gn,Beetz:2008ij,Federrath:2008ey,PriceFederrathBrunt2011,BurkhartLazarian2012,Seon2012,Konstandin:2012vy}, with investigation into different simulation techniques \citep{PriceFederrath2010} and stirring methods \citep{Federrath:2008ey,Federrath:2010ef}. It has also been studied by employing a heating and cooling curve \citep{WadaNorman2001,KritsukNorman2002,AuditHennebelle2005,AuditHennebelle2010,HennebelleAudit2007,SeifriedEtAl2011,GazolKim2013}. Recently an investigation has been done in the magnetohydrodynamic (MHD) regime \citep{Molina:2012iv}, and on polytropic gases \citep[][]{FederrathBanerjee2015}.

In our work, we investigate the robustness of this well-established density variance -- Mach number relation, Equation~(\ref{relation1}), in the non-isothermal regime, specifically in ideal gases with diatomic molecular ($\gamma = 7/5$) and monatomic ($\gamma = 5/3$) adiabatic indices.

This is relevant because theoretical models of the star formation rate \citep{KrumholzMcKee2005,PadoanNordlund2011,HennebelleChabrier2011,FederrathKlessen2012}, the star formation law \citep{Federrath2013sflaw}, the star formation efficiency \citep{Elmegreen2008}, and the initial mass function of stars \citep{HennebelleChabrier2008,Hopkins2013IMF,ChabrierEtAl2014} heavily rely on Equation~(\ref{relation1}).

Section~\ref{sec:methods} summaries our simulation and analysis methods, Section~\ref{sec:results} first presents results for the isothermal case, in order to make contact with previous studies and to verify our analysis techniques. Then we present a numerical resolution study to determine the minimum resolution required in order to measure the density variance -- Mach number relation in simulations with $\gamma>1$ and present our main results for adiabatic indices $\gamma=7/5$ and $5/3$. We provide a theoretical model for the $\sigma_s^2(\mathcal{M})$ relation in Section~\ref{sec:discussion} and discuss the discrepancies that we find compared to the standard Equation~(\ref{relation1}). Section~\ref{sec:conclusion} summarises our conclusions.

\section{Simulation and analysis methods} \label{sec:methods}

\begin{table*}
\caption{Simulation parameters. \textbf{Notes.} Column 1: simulation name. Column 2: grid resolution. Column 3: adiabatic exponent $\gamma$ in Equation~(\ref{eq:eos}). Column 4: dimensionless driving amplitude of the turbulence. Column 5: Resulting time-averaged velocity dispersion in code units in the regime of fully developed turbulence. Column 6: turbulent box crossing time: $t_\mathrm{cross}=L/\sigma_v$ in code units.}
\begin{tabular}{lrrrrr} 
\hline
Simulation name & $N^3_\mathrm{res}$ & $\gamma$ & $A$ & $\sigma_{v}$ & $t_\mathrm{cross}$ \\
\hline
(01) AD-TURB-256-A200-G1 & $256^3$ & 1.0001 & 200 & $2.19 \pm 0.14$ & $0.46\pm0.03$ \\
(02) AD-TURB-256-A400-G1 & $256^3$ & 1.0001 & 400 & $3.38 \pm 0.13$ & $0.30\pm0.01$\\
(03) AD-TURB-256-A800-G1 & $256^3$ & 1.0001 & 800 & $5.45 \pm 0.29$ & $0.18\pm0.01$ \\
(04) AD-TURB-256-A1600-G1 & $256^3$ & 1.0001 & 1600 & $8.62 \pm 0.61$ & $0.12\pm0.01$ \\
\hline
(05) AD-TURB-256-A100-G7/5 & $256^3$ & 7/5 & 100 & $1.53 \pm 0.04$ & $0.65\pm0.02$ \\
(06) AD-TURB-256-A200-G7/5 & $256^3$ & 7/5 & 200 & $2.47 \pm 0.12$ & $0.40\pm0.02$ \\
(07) AD-TURB-256-A400-G7/5 & $256^3$ & 7/5 & 400 & $4.00 \pm 0.17$ & $0.25\pm0.01$ \\
\hline		
(08) AD-TURB-1024-A200-G5/3 & $1024^3$ & 5/3 & 200 & $2.51 \pm 0.14$ & $0.40\pm0.02$ \\
(09) AD-TURB-512-A200-G5/3 & $512^3$ & 5/3 & 200 & $2.50 \pm 0.13$ & $0.40\pm0.02$ \\
(10) AD-TURB-256-A100-G5/3 & $256^3$ & 5/3 & 100 & $1.55 \pm 0.05$ & $0.65\pm0.02$ \\
(11) AD-TURB-256-A200-G5/3 & $256^3$ & 5/3 & 200 & $2.55 \pm 0.14$ & $0.39\pm0.02$ \\
(12) AD-TURB-256-A400-G5/3 & $256^3$ & 5/3 & 400 & $4.07 \pm 0.24$ & $0.25\pm0.01$ \\
(13) AD-TURB-128-A200-G5/3 & $128^3$ & 5/3 & 200 & $2.47 \pm 0.13$ & $0.40\pm0.02$ \\
(14) AD-TURB-64-A200-G5/3 & $64^3$ & 5/3 & 200 & $2.45 \pm 0.14$ & $0.41\pm0.02$ \\
\hline
\label{tab:sim}
\end{tabular}
\end{table*}

To simulate the turbulent ISM we use the high-resolution, shock-capturing code Fyris Alpha \citep{Sutherland:2010cp} to solve the equations of compressible hydrodynamics across a three-dimensional, periodic domain with side length $L = 1$, initial uniform density $\bar{\rho} = 1$, pressure of 1/2 ($c_{\rm s}^2 = \gamma/2$), and zero initial velocities. Unlike previous studies, the goal here is to test the density and velocity statistics of purely adiabatic turbulence with an ideal gas EOS, rather than a purely isothermal or polytropic EOS, and simpler than employing a cooling curve or running chemo-hydrodynamical simulations \citep{GloverFederrathMacLowKlessen2010}. Ultimately, simulations with multiple species including all relevant chemical reactions, as well as radiative heating and cooling would be the most realistic, but their complexity might not allow us to reduce the results to some simple rules of thumb that can be used in practical applications. We thus simplify the problem significantly by studying the turbulence in purely adiabatic, ideal gases with the aim of extracting results that might be applicable to a wider range of cases, including terrestrial experiments and atmospheric turbulence, in addition to the ISM. Table~\ref{tab:sim} lists the key parameters of all our adiabatic turbulence simulations.

\subsection{Ideal gas equation of state}

The ideal gas equation of state relates the pressure $P$, density $\rho$ and temperature $T$, and is given by
\begin{align} \label{eq:idealgas}
\frac{PV}{T}=N k_\mathrm{B} \quad\text{or}\quad \frac{P}{\rho}=\frac{N}{m}k_\mathrm{B}T \quad\text{or}\quad P=nk_\mathrm{B}T,
\end{align}
with the volume $V$, the Boltzmann constant $k_\mathrm{B}$, the particle mass $m$, the total number of particles $N$, and the particle number density $n=N/V$.
The ratio of the specific heat capacities at constant pressure and constant volume defines the adiabatic index,
\begin{align} \label{eq:gamma}
\gamma=\frac{c_P}{c_V}=1\,+\frac{2}{f},
\end{align}
where $f$ denotes the number of degrees of freedom. For monatomic gas, $f=3$ and $\gamma=5/3$, while for diatomic molecular gas, $f=5$ and $\gamma=7/5$, because diatomic molecules have two rotational degrees of freedom in addition to the three translational degrees of freedom. Note that at typical molecular cloud temperatures (about $10$--$100\,\mathrm{K}$), oscillatory degrees of freedom cannot be excited by collisions, which is why---although theoretically present---they do not contribute to increase $f$ in such cases. The specific internal energy of an ideal gas, $u=\frac{f}{2}\frac{N}{m}k_\mathrm{B}T$, is only determined by its temperature. Inserting this equation into Equations~(\ref{eq:idealgas}) and~(\ref{eq:gamma}), leads to
\begin{align} \label{eq:eos}
P(\rho,T)=(\gamma-1)\rho u(T)
\end{align}
expressed via the adiabatic index $\gamma$, which serves as the equation of state. In order to determine the statistics of turbulence in this adiabatic regime, we use isothermal, diatomic molecular and monatomic equations of state ($\gamma\to1, \gamma = 7/5$ and $5/3$ respectively).

Note that in order to model isothermal gases, $\gamma$ is often set close to unity (e.g., $\gamma=1.0001$), as if the gas had an extremely large number of degrees of freedom $f\to\infty$. This trick produces a gas that approximately stays at constant temperature, because any excess heat from dissipation (e.g., by shocks) is absorbed in such a big internal energy reservoir that the temperature of the gas does not notably change.

\subsection{Driving of turbulence, time evolution, and definition of the Mach number}

The driving of turbulence in the gas is performed by stirring with random purely solenoidal (divergence-free) forcing at low wavenumbers for the duration of the simulation. All wavenumbers $k$ in the range $1\leq k/(2\pi/L)\leq3$ were driven. The driving pattern is evolved with an Ornstein-Uhlenbeck process similar to the methods explained in \citet{EswaranPope1988}, \citet{SchmidtEtAl2009}, and \citet{FederrathDuvalKlessenSchmidtMacLow2010}.

From the work done by \citet{Federrath:2008ey,Federrath:2010ef}, we expect a proportionality constant of $b \sim 1/3$ in Equation~(\ref{relation1}) for solenoidally driven isothermal gas and we will test that in both isothermal and adiabatic gases. The rms Mach number of the gas is modified by varying the stirring amplitude $A$ of the driving force, allowing each $\gamma$ to be tested at a range of Mach numbers.

\begin{figure}
\centerline{\includegraphics[width=1.0\linewidth]{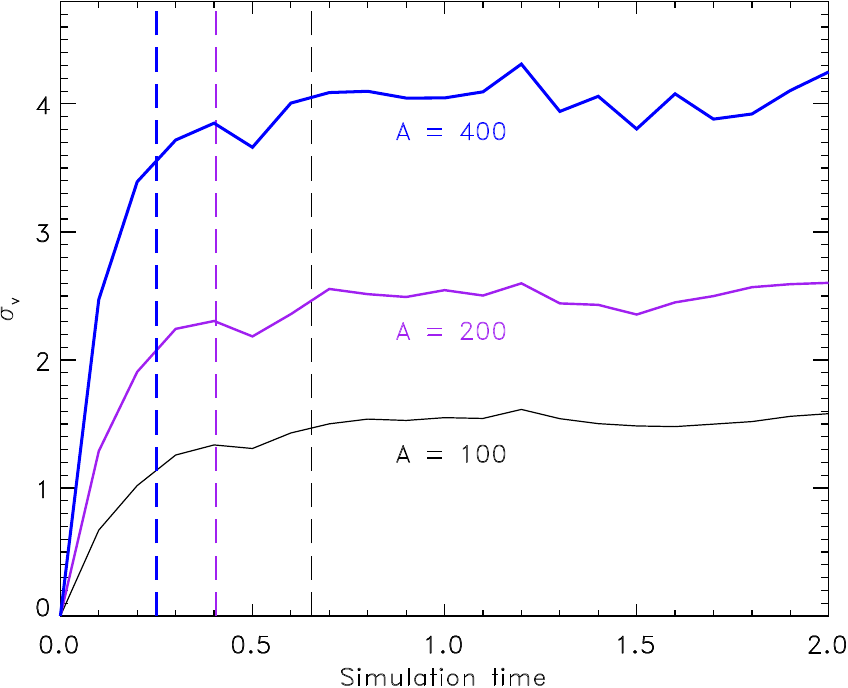}}
\caption{The velocity dispersion $\sigma_{v}$, as a function of simulation time for driving amplitudes $A = 100$, $200$ and $400$, and for fixed adiabatic $\gamma = 7/5$. The times for the onset of turbulence in each case are shown as vertical dashed lines and are approximated with the box crossing time $t_\mathrm{cross}=L/\sigma_v$, where $L$ is the linear size of the computational domain. The box crossing time for each simulation is listed in Table~\ref{tab:sim}.}
\label{amp1}
\end{figure}

All simulations are run for several turbulent box crossing times to test the density variance --  Mach number relation in the regimes of transient as well as fully-developed turbulence. The turbulent crossing time is defined as $t_\mathrm{cross} \equiv L/\sigma_v$, where $\sigma_v$ is the asymptotic velocity dispersion. The velocity dispersion and hence the crossing time depend on the driving amplitude $A$. We show an example of this dependence in Fig.~\ref{amp1}. We assume that the turbulence becomes fully developed after one crossing time in each simulation, indicated by the dashed vertical lines in Fig.~\ref{amp1}. At this point the gas properties no longer vary drastically but change smoothly, indicating a statistically stable configuration. This allows us to distinguish regimes of transient ($t<t_\mathrm{cross}$) and fully-developed ($t>t_\mathrm{cross}$) turbulence.

Given the velocity dispersion and sound speed $c_{\rm s}=(\partial P/\partial \rho)^{1/2}$, we have different Mach numbers $\mathcal{M}=\sigma_v/c_{\rm s}$, depending on the driving amplitude $A$ and depending on the value of adiabatic $\gamma$. This is because the sound speed depends on the derivative of the EOS, Equation~(\ref{eq:eos}). It furthermore depends on the simulation time, because the internal energy and thus the mean temperature of the gas keeps increasing during the course of the adiabatic simulations with $\gamma=7/5$ and $\gamma=5/3$. This is in stark contrast to the isothermal and polytropic simulations performed in previous studies \citep{Padoan:1997tn,Passot:1998wz,Federrath:2008ey,PriceFederrathBrunt2011,Konstandin:2012vy,FederrathBanerjee2015}, where the sound speed did not change systematically, after the turbulence was fully developed. Here, however, the total energy is conserved, which means that all dissipated energy is conservatively added to the internal energy. Thus, the injected energy from the driving is converted into internal energy and heats the gas continuously, leading to an ever increasing average sound speed and to a continuously decreasing rms Mach number. We thus use instantaneous measurements of $\mathcal{M}$ and $\sigma_s$ in the following to determine the density variance -- Mach number relation in adiabatic gases.

\subsection{Measuring the density variance}

The density variance of the gas is calculated using method 4 in Section 2.3 of \citet{PriceFederrathBrunt2011}, but instead of fitting a lognormal distribution, we fit the more appropriate \citet{Hopkins2013PDF} distribution. The advantage of the Hopkins fit is that it takes turbulent intermittency effects into account and provides excellent fits to the density PDFs over a wide range of physical parameters, including different Mach numbers, driving amplitudes and mixtures \citep{Federrath2013}, as well as magnetic field strengths and variations in the polytropic exponent  for simulations that employ a polytropic EOS \citep{FederrathBanerjee2015}. The \citet{Hopkins2013PDF} density PDF is defined as
\begin{align} \label{eq:hopkinspdf}
p_\mathrm{HK}(s) = I_1\left(2\sqrt{\lambda\,\omega(s)}\right)\exp\left[-\left(\lambda+\omega(s)\right)\right]\sqrt{\frac{\lambda}{\theta^2\,\omega(s)}}\,,\nonumber\\
\lambda\equiv\sigma_s^2/(2 \theta^2),\quad\omega(s)\equiv\lambda/(1+ \theta)-s/ \theta \;\; (\omega\geq0),
\end{align}
where $I_1(x)$ is the modified Bessel function of the first kind. Equation~(\ref{eq:hopkinspdf}) is motivated and explained in detail in \citet{Hopkins2013PDF}. It contains two parameters: 1) the volume-weighted standard deviation of logarithmic density fluctuations $\sigma_s$, and 2) the intermittency parameter $ \theta$. In the zero-intermittency limit ($\theta\to0$), Equation~(\ref{eq:hopkinspdf}) becomes the lognormal distribution from Equation~(\ref{pdf}), $p_\mathrm{HK}\to p_\mathrm{LN}$.

In order to measure the density variance $\sigma_s^2$, we fit our simulation density PDFs in a restricted range around the mean (from $\bar{s}-3\sigma_{s\rm, mom}$ to $\bar{s}+3\sigma_{s\rm,mom}$, where $\sigma_{s\rm,mom}$ is the second moment of the density distribution, directly computed by summation over all simulation data points) with Equation~(\ref{eq:hopkinspdf}) and determine the best-fit parameter $\sigma_s$. In agreement with the conclusions drawn in \citet{PriceFederrathBrunt2011} and \citet{Hopkins2013PDF}, we find that the fitted $\sigma_s$ is the same within a few percent as $\sigma_{s\rm,mom}$ (computed by summation over all simulation grid cells).

\section{Results} \label{sec:results}

\subsection{Isothermal comparison} \label{sec:iso}

\begin{figure}
\centerline{\includegraphics[width=1.0\linewidth]{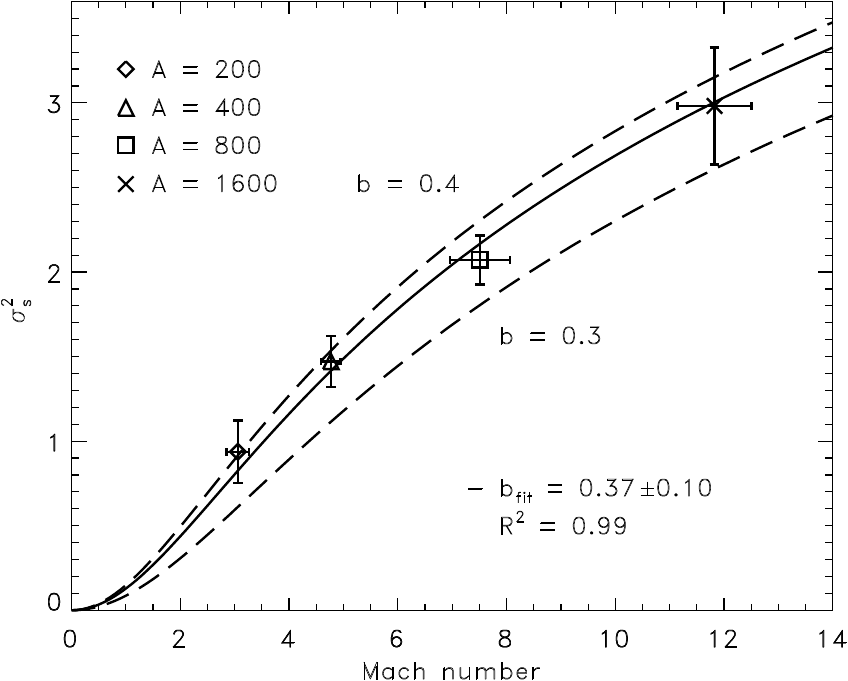}}
\caption{The density variance -- Mach number relation for isothermal turbulence (approximated by setting $\gamma = 1.0001$). In order to reach a wide range of Mach numbers to test the relation, we use stirring amplitudes between $A = 200$ and $1600$ (\mbox{models~1--4} in Table~\ref{tab:sim}), leading to Mach numbers between 3 and 12. We fit Equation~(\ref{relation1}) to the four points and find the proportionality constant $b = 0.37 \pm 0.10$ with a goodness of fit of R$^2 = 0.99$. The fit is shown as a solid line, while cases with $b=0.3$ and $b=0.4$ are shown as dashed lines for comparison. Our best fit is consistent with the expectation value $b\sim1/3$ for purely solenoidal driving \citep{Federrath:2008ey,Federrath:2010ef}.}
\label{fig:Iso1}
\end{figure}

The isothermal gas relation has been studied extensively and is therefore a good comparison test for our hydrodynamics code, setup and post-processing methods to determine $\sigma_s$ (see~\S\ref{sec:methods}). For purely solenoidal forcing of the turbulence, we expect a proportionality value of $b\sim1/3$ in Equation~(\ref{relation1}) \citep{Federrath:2008ey,Federrath:2010ef}. Four simulations were performed at a grid resolution of $256^3$ with $\gamma = 1.0001$ to prevent non-isothermal effects brought on by high stirring amplitudes.  The four time-averaged points lie between $b = 0.3$ and 0.4, with a fit of $b = 0.37 \pm 0.10$ and goodness-of-fit parameter R$^2 = 0.99$ (Fig. \ref{fig:Iso1}). Our measurement of $b$ spans the expected value, thus our methods produce reasonable results for isothermal turbulence. Now that we have established that our simulation and analysis techniques reproduce previous results for isothermal turbulence, we can now move on to study non-isothermal turbulence in the adiabatic regime.

\subsection{Resolution study} \label{sec:res}

\begin{figure*}
\centerline{\includegraphics[width=1.0\linewidth]{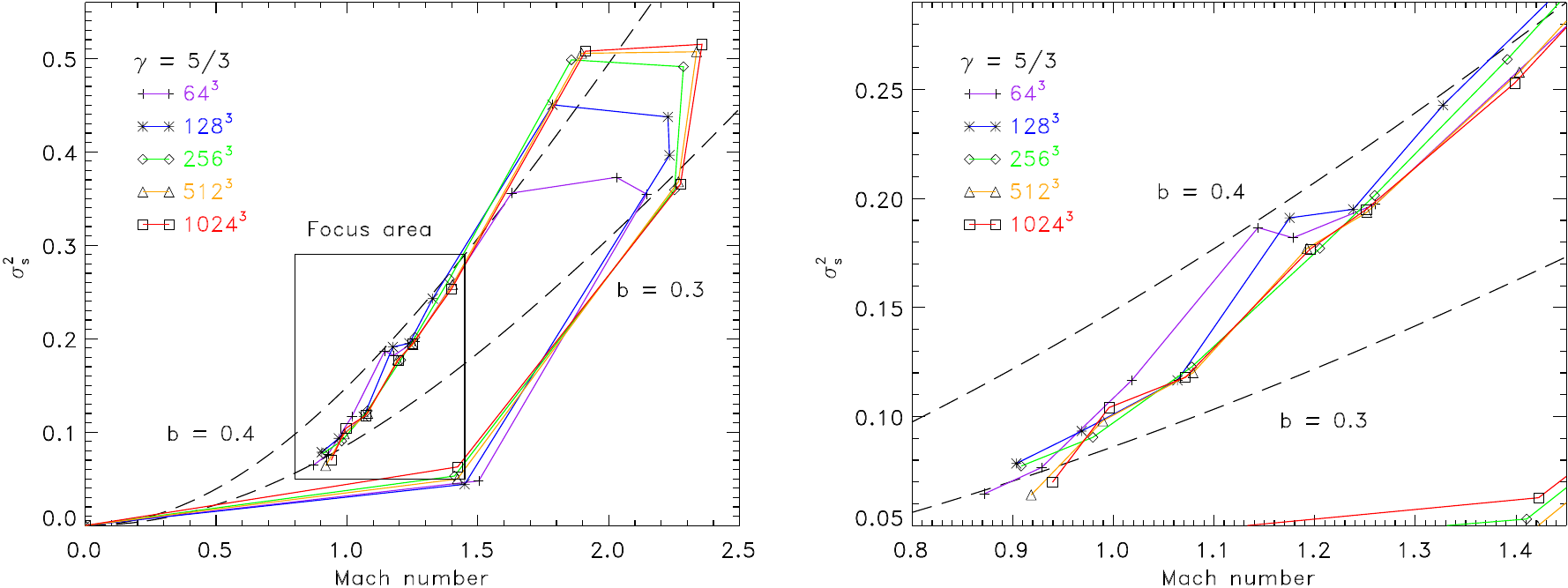}}
\caption{Left panel: Evolution of simulations with increasing resolution in density variance -- rms Mach number space. The dashed lines represent functions of Equation~(\ref{relation1}) with $b = 0.3$ and $0.4$, for comparison. Initially the rms Mach number increases substantially, then decreases smoothly once the turbulence becomes fully developed (at about $t_\mathrm{cross}$), as a result of the continuously increasing internal energy, temperature and sound speed for our ideal gas EOS, Equation~(\ref{eq:eos}). Right panel: A magnified region of the left panel, displaying the convergence of statistics. Simulations with $\ge 256^3$ grid cells are representative of the converged system.}
\label{fig:res}
\end{figure*}

We now test the resolution requirements of density variance and Mach number measurements by performing a series of identical simulations at increasing resolutions of $64^3$, $128^3$, $256^3$, $512^3$ and $1024^3$ grid cells for $\gamma = 5/3$. Fig.~\ref{fig:res} shows the density variance -- Mach number relation for each simulation and simulation time. We see that first, the density variance and Mach number increase and reach a maximum after about $t_\mathrm{cross}$ (the lower part of the correlation). In this first part of the evolution, the kinetic energy of the gas increases due to the driving until the kinetic energy power spectrum is established \citep{SchmidtEtAl2009}. After the kinetic energy and rms velocity have reached a saturated state (see Fig.~\ref{amp1}), only the sound speed keeps increasing monotonically, because the dissipated energy heats the gas. This leads to a continuously decreasing Mach number and density variance in the regime of fully developed turbulence (the upper part of the correlation).

The higher the numerical grid resolution, the greater the maximum of $\sigma_s^2$ for $N_\mathrm{res}<256$, which is seen to converge for $N_\mathrm{res}\gtrsim256$. A zoom within the focus area of the density variance -- Mach number relation is shown in the right-hand panel of Fig.~\ref{fig:res}. These points are independent of the initial jump, but are still seen to converge at increased resolution.  For resolutions equal to and above $256^3$, the values are almost identical. Therefore the statistics of density variance and rms Mach number may be approximated well by numerical resolutions $\ge 256^3$ cells. This is consistent with the resolution requirements established in \citet{Kitsionas:2009di}, \citet{Federrath:2010ef}, \citet{KritsukEtAl2011codes}, and \citet{Federrath2013}.

Looking closely at Fig.~\ref{fig:res}, we see that the gas evolves along a curve in the density variance -- Mach number plane. This is due to the continuous heating of the gas via the turbulent driving, which lowers the Mach number continuously. The evolution of this curve seems to correlate somewhat with Equation~(\ref{relation1}), but is steeper than the theoretical prediction for isothermal turbulence. The behaviour of this curve is quantified in detail in the following subsections, \S\ref{sec:case1} and~\S\ref{sec:case2} for $\gamma = 7/5$ and $\gamma = 5/3$, respectively.

\subsection{Diatomic molecular gas: $\gamma = 7/5$} 
\label{sec:case1}

\begin{figure}
\centerline{\includegraphics[width=1.0\linewidth]{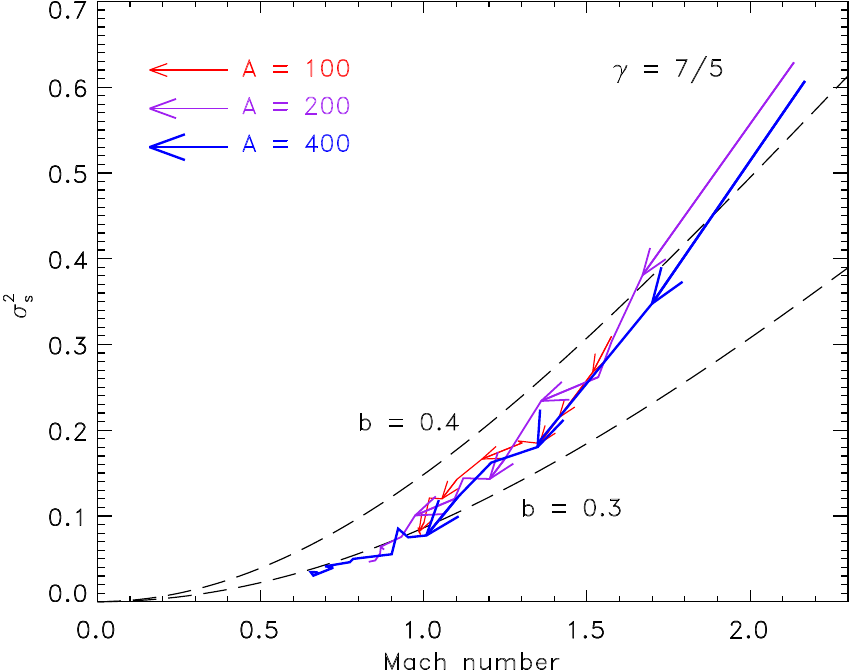}}
\caption{Evolution of simulations with $\gamma = 7/5$ and $A = 100$ (red, small arrows), 200 (purple, normal arrows) and 400 (blue, large arrows). The arrows indicate the direction of time evolution.}
\label{fig:g1.4}
\end{figure}

In the first case we look at a diatomic equation of state, i.e., $\gamma=7/5$. We perform three separate simulations with different stirring amplitudes $A$, and plot their evolutionary curves in Fig.~\ref{fig:g1.4}. 
The best fit to Equation~(\ref{relation1}) from the combined points of all three simulations is $b = 0.37 \pm 0.02$ with a goodness of fit parameter of R$^2 = 0.90$. Note that the increase in sound speed due to gas heating is slow compared to the time it takes to establish a statistically steady state, which is shown in Fig.~\ref{amp1}, where we see that the velocity dispersion is fully developed after one crossing time. However, the continuous heating in the fully developed regime of turbulence leads to a continuously increasing sound speed, the consequences of which are discussed in more detail in Section~\ref{sec:discussion}.

As in Fig.~\ref{fig:res}, we see in Fig.~\ref{fig:g1.4} that simulations with $\gamma>1$ produce a somewhat steeper $\sigma_s^2(\mathcal{M})$ relation compared to the isothermal relation (cf.~Fig.~\ref{fig:Iso1}) as time progresses and both $\sigma_s^2$ and $\mathcal{M}$ decrease due to the continuous heating of the gas. The different times of each simulation are connected by a line with arrows in Fig.~\ref{fig:g1.4}, indicating increasing time. Therefore a new functional fit might be more appropriate to describe the behaviour in our non-isothermal, adiabatic turbulence simulations.

The simplest modification to the existing model function, Equation~(\ref{relation1}), is to allow for variations in the exponent on the Mach number. Our data in Fig.~\ref{fig:g1.4} indicate that the exponent is somewhat higher than the standard $\mathcal{M}^2$ dependence from Equation~(\ref{relation1}). Thus we use the following new fit function to determine the exponent $\alpha$:
\begin{equation} \label{eq:newfit}
\sigma^2_{s} = \ln(1 + b^{\prime2} \mathcal{M}^{\alpha}).
\end{equation}
We do not necessarily expect that the coefficient $b^{\prime}$ in this new relation is the same as $b$ in Equation~(\ref{relation1}), but we will test that below. 
The new function is fitted to the data from each of the three simulations with different driving amplitude and to the combined set of data points. We determine the goodness of fit parameter $R^2$ for the fits to Equations~(\ref{relation1}) and~(\ref{eq:newfit}) and compare them. The results are summarised in Table~\ref{y140_fits}.

\begin{table*}
\centering
\caption{Statistical fit parameters for different functions $\sigma_s^2(\mathcal{M})$, for $\gamma = 7/5$. \textbf{Notes.} $R^2$ denotes the goodness-of-fit parameter. A value of $R^2=1$ indicates a perfect fit to the given data.}
\begin{tabular}{cccccc} 
\hline
Fit function & \multicolumn{2}{c}{$\sigma^2_{s} = \ln(1 + b^2 \mathcal{M}^2)$} & \multicolumn{3}{c}{$\sigma^2_{s} = \ln(1 + b^{\prime2} \mathcal{M}^{\alpha})$} \\
Parameter & $b$ & $R^2$ & $b^{\prime}$ & $\alpha$ & $R^2$ \\\hline
$A = 100$ & $0.35\pm0.02$ & 0.90 & $0.32\pm0.05$ & $2.7\pm0.8$ & 0.96 \\
$A = 200$ & $0.38\pm0.04$ & 0.90 & $0.31\pm0.07$ & $2.9\pm0.8$ & 0.99 \\
$A = 400$ & $0.37\pm0.04$ & 0.92 & $0.30\pm0.07$ & $2.8\pm0.8$ & 1.0 \\
All data     & $0.37\pm0.02$ & 0.90 & $0.31\pm0.04$ & $2.8\pm0.4$ & 0.99 \\
\hline
\label{y140_fits}
\end{tabular}
\end{table*}

Table~\ref{y140_fits} shows that for the individual simulations as well as for the combined data set, the modified power-law function from Equation~(\ref{eq:newfit}) provides the better fit to the data as quantified by the goodness of fit $R^2$. The coefficient value $b^{\prime}=0.31\pm0.04$ is smaller than $b=0.37\pm0.02$, but they are formally consistent with representing the same value, and consistent with the $b$-value obtained in our isothermal calculations in~\S\ref{sec:iso}. In fact, our new fit gives a value that is in agreement with the theoretical expectation for the turbulent driving, namely $b\sim1/3$ for purely solenoidal driving as applied here.

The exponent $\alpha$, which is fixed to $\alpha=2$ in Equation~(\ref{relation1}), but allowed to vary in our new fit function, Equation~(\ref{eq:newfit}), clearly shows that an almost cubic dependence on $\mathcal{M}$ provides a better fit to the data. We find a best-fit value of $\alpha=2.8\pm0.4$ in our simulations with $\gamma=7/5$, leading to a new form of the density variance -- Mach number relation for $\gamma=7/5$ gas,
\begin{equation}
\sigma^2_{s} = \ln\left[ 1+ (0.31 \pm 0.04)^2 \cdot \mathcal{M}^{(2.8 \pm 0.4)} \right].
\end{equation}

This result presents an interesting question: is the density variance -- rms Mach number relation of non-isothermal adiabatic turbulence no longer a quadratic relation, compared to the well-studied isothermal case? We will now explore whether the same/similar holds for $\gamma=5/3$ and then address this questions in the discussion of Section~\ref{sec:discussion}, by comparing to a theoretical model of the $\sigma_s^2(\mathcal{M})$ relation.

\subsection{Monatomic gas: $\gamma = 5/3$} \label{sec:case2}

\begin{figure}
\centerline{\includegraphics[width=1.0\linewidth]{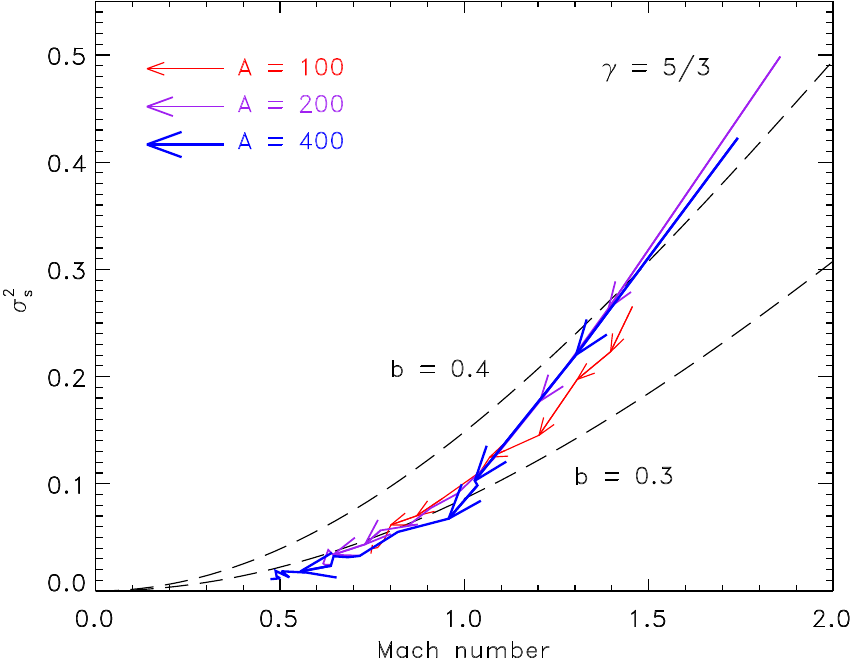}}
\caption{Evolution of simulations with $\gamma = 5/3$ and $A = 100$ (red, small arrows), 200 (purple, normal arrows) and 400 (blue, large arrows). The arrows indicate the direction of time evolution.}
\label{g1.66}
\end{figure}

In the second case we look at a monatomic EOS, Equation~(\ref{eq:eos}) with $\gamma=5/3$. As in the diatomic case we performed three separate simulations with different driving amplitude $A$, and plot their evolutionary curves in Fig.~\ref{g1.66} (again with arrows indicating the continuous time evolution to smaller and smaller $\sigma_s^2$ and $\mathcal{M}$). A fit with the standard relation, Equation~(\ref{relation1}), to the combined data set gives $b = 0.36 \pm 0.02$ with a goodness of fit $R^2=0.90$.

Given the same effect occurs as for $\gamma = 7/5$, we fit our new power-law function, Equation~(\ref{eq:newfit}), to data from each of the three simulations and to the combined set of data points, and compare the goodness of fit values with those from the fit to Equation~(\ref{relation1}). The results are summarised in~Table \ref{y166_fits}.

\begin{table*}
\centering
\caption{Statistical fit parameters for different functions $\sigma_s^2(\mathcal{M})$, for $\gamma = 5/3$. \textbf{Notes.} $R^2$ denotes the goodness-of-fit parameter. A value of $R^2=1$ indicates a perfect fit to the given data.}
\begin{tabular}{cccccc} 
\hline
Fit function & \multicolumn{2}{c}{$\sigma^2_{s} = \ln(1 + b^2 \mathcal{M}^2)$} & \multicolumn{3}{c}{$\sigma^2_{s} = \ln(1 + b^{\prime2} \mathcal{M}^{\alpha})$} \\
Parameter & $b$ & $R^2$ & $b^{\prime}$ & $\alpha$ & $R^2$ \\\hline
$A = 100$ & $0.34\pm0.03$ & 0.92 & $0.32\pm0.05$ & $2.8\pm0.9$ & 0.99 \\
$A = 200$ & $0.38\pm0.04$ & 0.91 & $0.33\pm0.06$ & $2.9\pm0.8$ & 1.0 \\
$A = 400$ & $0.36\pm0.04$ & 0.89 & $0.32\pm0.06$ & $3.0\pm0.9$ & 1.0 \\
All data     & $0.36\pm0.02$ & 0.90 & $0.32\pm0.03$ & $3.0\pm0.5$ & 0.99 \\
\hline
\label{y166_fits}
\end{tabular}
\end{table*}

Once again we see a significantly better fit to the power law with exponent $\alpha>2$. We find that the driving coefficient $b^{\prime}\sim b$, as for the $\gamma=7/5$ case, indicating that the physics of the driving is indeed contained in the value of the $b$ parameter, while the fact that we deal with non-isothermal turbulence is reflected in a steeper power-law exponent $\alpha>2$, compared to the isothermal case. All three simulations fit exponents very close to cubic ($\alpha\sim3$), with the combined data for $\gamma=5/3$ fitting a power law of the form
\begin{equation}
\sigma^2_{s} = \ln\left[ 1+ (0.32 \pm 0.03)^2 \cdot \mathcal{M}^{(3.0 \pm 0.5)} \right].
\end{equation}

\begin{figure}
\centerline{\includegraphics[width=1.0\linewidth]{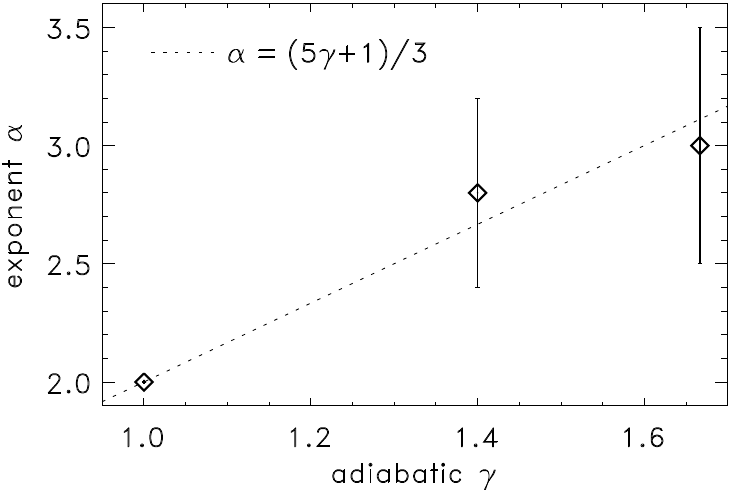}}
\caption{Exponent $\alpha$ in the density variance -- Mach number relation for different values of the adiabatic index $\gamma$. The data points show our simulation measurements and the dotted line is a linear fit with $\alpha=(5\gamma+1)/3$.}
\label{fig:alpha}
\end{figure}

\subsection{Summary of the results} \label{sec:sumad}

In summary, both the $\gamma=7/5$ and $\gamma=5/3$ cases yield turbulent driving coefficients $b\sim1/3$ that are all consistent with the theoretical expectation for purely solenoidal driving of the turbulence, and consistent with the $b$-values found for isothermal turbulence ($\gamma\to1$).

The exponent $\alpha$ of the $\sigma_s^2(\mathcal{M})$ relation, however, is significantly steeper with $\alpha \sim 2.8\pm0.4$ for $\gamma=7/5$ and $\alpha=3.0\pm0.5$ for $\gamma=5/3$, compared to the isothermal case, where $\alpha=2$ provides the best fit to the data. Thus, we see that the exponent $\alpha$ in the density variance -- Mach number relation is $\gamma$-dependent.

In order to provide a heuristic relation that describes the behaviour of $\sigma_s^2(\mathcal{M})$ in adiabatic gases, we fit a linear function to the dependence of $\alpha$ on $\gamma$ in Fig.~\ref{fig:alpha}. It is reasonable that $\alpha$ will continuously increase with $\gamma$. Thus, we chose the simplest function to approximate our data from $\gamma=1$ to $\gamma=5/3$, i.e., a linear interpolation. The result is $\alpha=(5\gamma+1)/3$. The actual dependence might be somewhat different, but to determine a better fit would require us to measure $\alpha$ for a range of $\gamma$ values, in much smaller steps $\Delta\gamma$. This is beyond the scope of the paper, but we can already provide a new improved functional form of the density variance -- Mach number relation that takes the adiabatic index $\gamma$ into account. The best-fit function is given by
\begin{equation} \label{eq:result}
\sigma^2_{s} = \ln\left[ 1+ b^2 \mathcal{M}^{(5\gamma+1)/3} \right],
\end{equation}
which is the central result of the paper. Equation~(\ref{eq:result}) naturally simplifies to the well-studied isothermal case ($\gamma\to1$) given by Equation~(\ref{relation1}), but also approximately covers cases with $\gamma>1$, up to $\gamma=5/3$.

\section{Discussion} \label{sec:discussion}

In \S\ref{sec:case1}--\S\ref{sec:sumad}, we found that the density variance -- Mach number relation for adiabatic gases with $\gamma=7/5$ and $5/3$ respectively, deviates significantly from the isothermal case, Equation~(\ref{relation1}). We quantified the discrepancy by fitting an alternative function, Equation~(\ref{eq:newfit}), to the data, with the power-law exponent $\alpha$ as a free fit parameter. We find that the power law provides excellent fits, with power-law exponents increasing with $\gamma$ from $\alpha=2$ for the isothermal case ($\gamma\to1$) to $\alpha\sim3$ for $\gamma=5/3$. A heuristic function was obtained to provide a new $\sigma_s^2(\mathcal{M})$ relation that takes the dependence on $\gamma$ into account, given by Equation~(\ref{eq:result}).

We now compare this results to a recent theoretical model for the density variance -- Mach number relation, in order to explain the differences of our adiabatic case to the isothermal and polytropic cases. The detailed derivation of the relation can be found in \citet{Molina:2012iv} and \citet{FederrathBanerjee2015}, where this relation has been explored for magnetized isothermal and polytropic gases respectively, with the latter representing a special case, where the pressure and temperature of the gas are both uniquely related to the density via
\begin{equation} \label{eq:polytropiceos}
P(\rho) \sim \rho^\Gamma, \quad T(\rho) \sim \rho^{\Gamma-1}.
\end{equation}
We emphasize that this is different from the adiabatic EOS, Equation~(\ref{eq:eos}), used here, in that the pressure depends on both density and temperature, $P(\rho,T)$. Thus, for any given value of density, the pressure can vary depending on the temperature, while a polytropic EOS will give only a single value of $P$ for a given input $\rho$.

The basic idea of the theoretical model is to relate the density jump in a single shock to the ensemble of shocks/compressions in a turbulent medium. For that purpose, \citet{PadoanNordlund2011} and \citet{Molina:2012iv} applied the equations of mass, momentum and energy conservation across a shock,
\begin{align}
\rho_0 v_{0} & = \rho v \label{eq:rh1}, \\
\rho_0 v_{0}^2 + P_0 & = \rho v^2 + P \label{eq:rh2}, \\
\frac{1}{2}v_0^2+u_0+\frac{P_0}{\rho_0} & = \frac{1}{2}v^2+u+\frac{P}{\rho} \label{eq:rh3},
\end{align}
to derive the density contrast $\rho/\rho_0$ between the pre-shock gas (denoted with index 0 and on the left-hand side of the equations) and the post-shock gas (no index; right-hand side of the shock jump equations). The equation of state, Equation~(\ref{eq:eos}), enters through the pressure $P$ and specific internal energy $u$ in these expressions. Combining these equations leads to the well-known Rankine-Hugoniot shock jump conditions as the solution for the density jump across the shock \citep{Rankine1870,Hugoniot1887,ShullDraine1987},
\begin{equation} \label{eq:shockjump}
\frac{\rho}{\rho_0} = \frac{v_0}{v} = \frac{(\gamma+1)b^2\mathcal{M}^2}{(\gamma-1)b^2\mathcal{M}^2+2}.
\end{equation}
Note that we have already introduced the geometrical $b$-parameter, because these shock jump conditions only apply to the plane-parallel component of the shock, parametrized by the parallel component of the sonic Mach number $v_0/c_{s,0}=b\mathcal{M}$ \citep{Molina:2012iv,FederrathBanerjee2015}. Following the detailed derivation in \citet{Molina:2012iv}, Equation~(\ref{eq:shockjump}) just needs to be inserted into the general expression for the ensemble of such shocks,
\begin{equation}
\sigma_s^2 = \ln\left( 1 + \frac{\rho}{\rho_0} \right),
\end{equation}
which leads to the density variance -- Mach number relation,
\begin{equation} \label{eq:theory}
\sigma_s^2 = \ln\left( 1 + \frac{(\gamma+1)b^2\mathcal{M}^2}{(\gamma-1)b^2\mathcal{M}^2+2} \right).
\end{equation}
We immediately see that this new $\gamma$-dependent density variance -- Mach number relation reduces to the isothermal case, Equation~(\ref{relation1}), if we set $\gamma=1$. In the adiabatic case, however, $\gamma>1$, which leads to our theoretical prediction as a function of $\gamma$, given by Equation~(\ref{eq:theory}).

\begin{figure*}
\centerline{\includegraphics[width=0.75\linewidth]{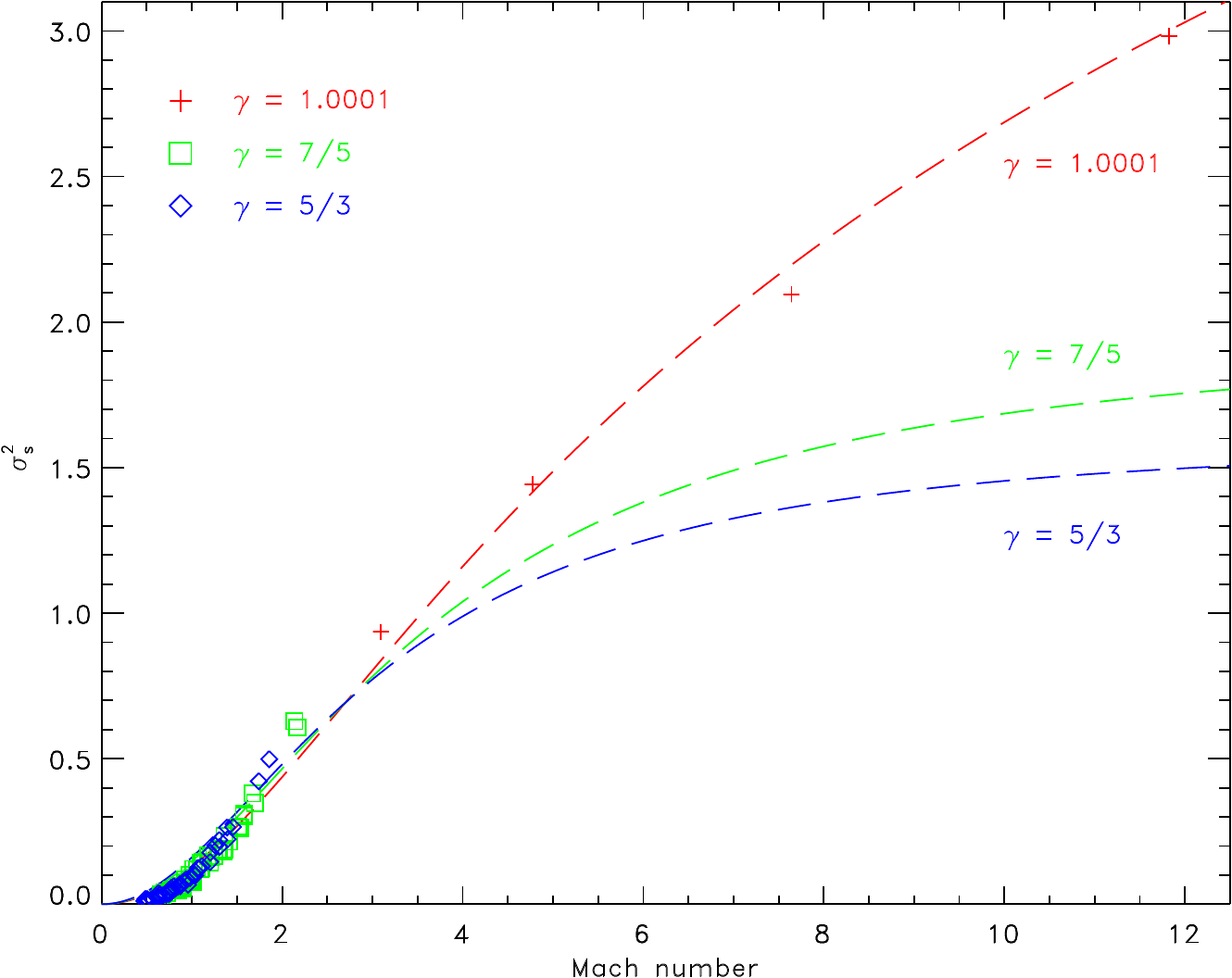}}
\caption{Combined density variance -- Mach number relation plot, showing all our simulation data with $\gamma=1.0001$ (red crosses), $\gamma=7/5$ (green boxes), and $\gamma=5/3$ (blue diamonds). The dashed lines are our theoretical prediction given by Equation~(\ref{eq:theory}) with the respective values of $\gamma$ (in the same colour as the simulation data and labelled on each curve).}
\label{fig:theory}
\end{figure*}

Fig.~\ref{fig:theory} shows the theoretical prediction given by Equation~(\ref{eq:theory}) for different values of $\gamma$ together with the isothermal solution and together with our simulation data for $\gamma=7/5$ and $\gamma=5/3$. We see that the new relation qualitatively follows the trend of a slightly steeper rise with increasing $\gamma$ for low Mach number. It also predicts that at high Mach number, the density variance saturates at lower $\sigma_s^2$ for increasing $\gamma$. This is reasonable, because the density jump across shocks reduces significantly with increasing $\gamma$ as derived in Equation~(\ref{eq:shockjump}). This regime needs to be tested in follow-up simulations that reach higher Mach numbers. However, the problem is that the adiabatic heating increases with increasing Mach, such that it quickly counteracts the effect of an increased driving amplitude.

Despite these reasonable qualitative trends produced by our new theoretical relation, Equation~(\ref{eq:theory}), we see that the actual simulation data with $\gamma>1$ still follow a somewhat steeper curve at low Mach number, $b\mathcal{M}\lesssim1$ in the $\sigma_s^2$--$\mathcal{M}$ plane. We speculate that this discrepancy arises, because the theoretical model does not contain any information about the temporal evolution of the gas, in particular about its temperature changes along the evolutionary curve.

However, we can qualitatively argue that any shock will immediately experience the temperature and pressure increase associated with the adiabatic compression. This will reduce the density jump significantly, such that the density variance will be smaller with increasing $\gamma$ almost immediately when these shocks are about to form, (e.g., the theoretical limit for $\gamma=5/3$ is $\rho/\rho_0=4$). Thus, shocks in high-$\gamma$ gas will be significantly reduced and so will be the statistical variance of density fluctuations, $\sigma_s^2$. The important point here is that this process almost instantaneously reduces the density variance.

At the same time, each shock dissipates energy, locally increasing the temperature and internal energy of the ideal adiabatic gas. This increases the sound speed, but at first only locally in the shocks, which leads to a decreasing pre-shock sonic Mach number over time, resulting in the time evolution (shown as arrows) in Figs.~\ref{fig:g1.4} and~\ref{g1.66}. We can thus qualitatively understand the time dependence and resulting $\sigma_s^2(\mathcal{M})$ relations for $\gamma>1$. The $\sigma_s^2(\mathcal{M})$ relation is steeper, because $\sigma_s^2$ responds almost instantaneously to the local pressure and temperature increase in shocks, while the Mach number reduction is delayed, because the sound speed increases only in the post-shock gas, while our theoretical Equation~(\ref{eq:theory}) is based on the large-scale, volume-weighted pre-shock Mach number.

\begin{figure*}
\centerline{\includegraphics[width=1.05\linewidth]{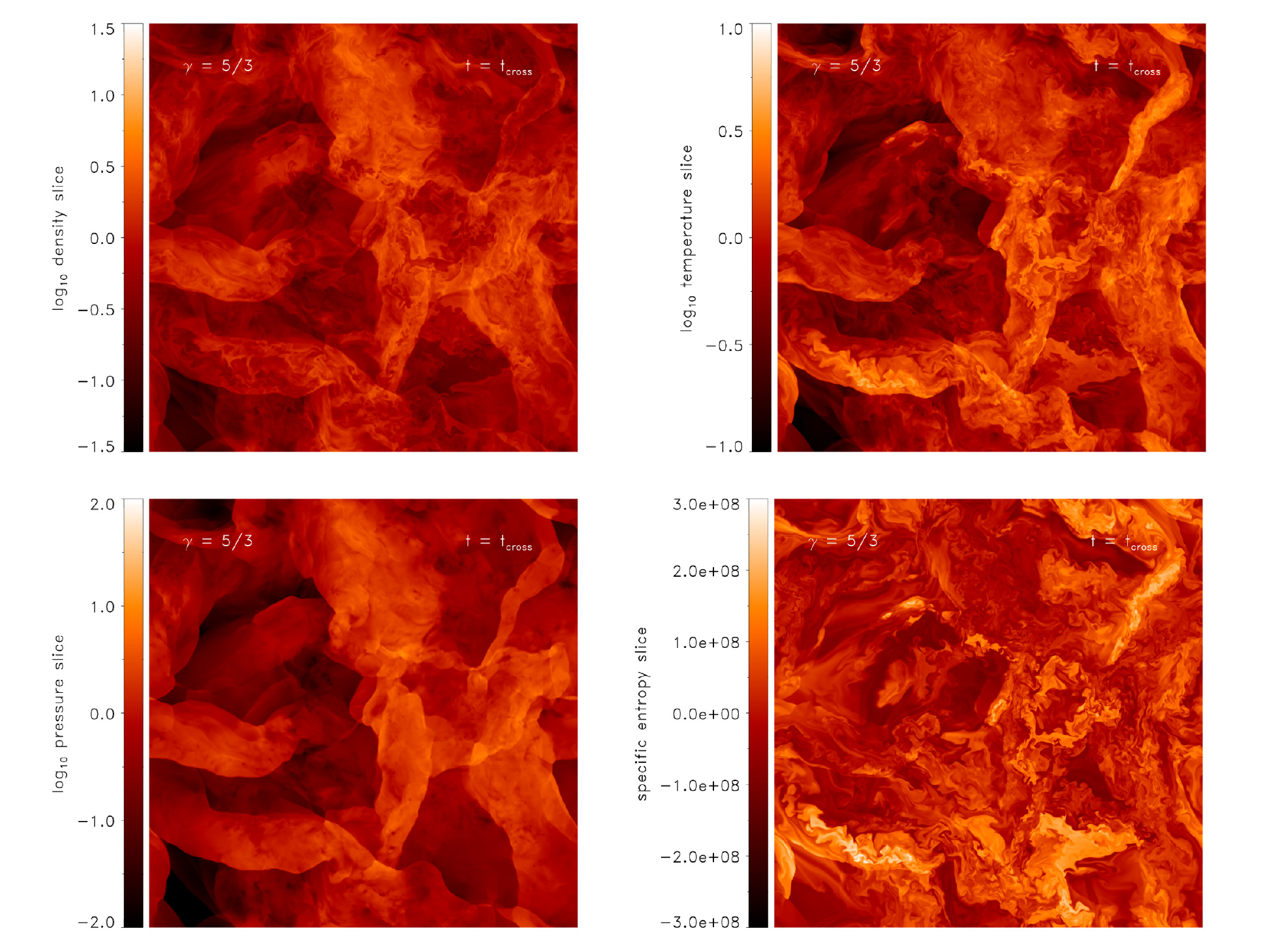}}
\caption{Slices of the normalized density (top left), temperature (top right), pressure (bottom left), and specific entropy (bottom right) at \mbox{$t=t_\mathrm{cross}$} for $\gamma = 5/3$, $A = 200$, and a numerical resolution of $1024^3$. It is evident that gas at a given density can have a wide range of temperatures and pressures, unlike a polytropic EOS where $T$ and $P$ are unique functions of the density only.}
\label{fig:slices}
\end{figure*}

In order to substantiate our findings, we show density, temperature, pressure, and entropy slices of our highest-resolution simulation with $1024^3$ grid cells and adiabatic $\gamma=5/3$ in Fig.~\ref{fig:slices}. We see two important points. First, the pressure and temperature are not unique functions of the density, but for a given density, the gas can have a range of temperatures and pressures, as implied by Equation~(\ref{eq:eos}). This is substantiated by the entropy slice shown in the bottom right panel of Figure~\ref{fig:slices}, which is not uniform, demonstrating that the gas is not barotropic, but that the pressure depends on density \emph{and} temperature. There is clearly viscous heating, which is primarily due to shocks in the $b\mathcal{M}>1$ regime, while turbulent dissipation (eddy viscosity) becomes a more important heating mechanism when $b\mathcal{M}<1$. Quantifying both contributions is beyond the scope of this paper. Second, the adiabatic heating primarily occurs in the post-shock gas. The rise of the internal energy does not immediately reduce the global post-shock Mach number, but slowly diffuses to large scales, before it affects $\mathcal{M}$, leading to the steeper-than-isothermal $\sigma_s^2(\mathcal{M})$ relations we found for $\gamma>1$.

\begin{figure*}
\centerline{\includegraphics[width=1.03\linewidth]{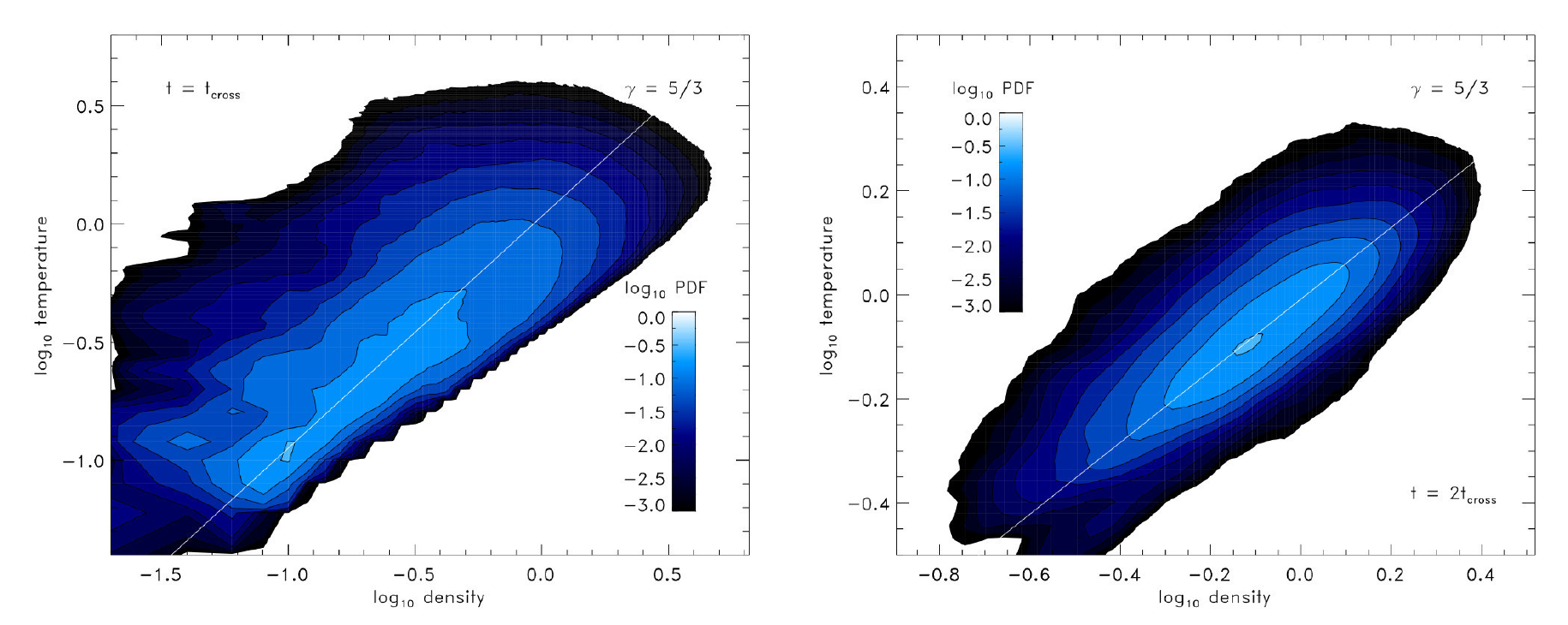}}
\caption{Density--temperature correlation PDFs for our adiabatic turbulence simulations with $\gamma=5/3$, $A=200$ and $N_\mathrm{res}=1024$. The left-hand panel shows the results at $t=t_\mathrm{cross}$, while the right-hand panel is for $t=2 t_\mathrm{cross}$. The distributions are spread by more than an order of magnitude for typical gas densities, around the average correlations (shown as white lines) with $T\sim\rho^{\gamma-1}$. The continuous heating of the gas indicated by the overall rise in temperature at $t=2\,t_\mathrm{cross}$ compared to $t=t_\mathrm{cross}$ primarily occurs in the post-shock gas, while the Mach number entering Equation~(\ref{eq:theory}) only applies to the pre-shock gas. This produces the steeper dependence of $\sigma_s^2$ on $\mathcal{M}$ that we find in our central result, Equation~(\ref{eq:result}).}
\label{fig:corrpdfs}
\end{figure*}

Finally, Fig.~\ref{fig:corrpdfs} shows density--temperature correlation probability density functions (PDFs). It is evident that for any given density, there is a wide range of temperatures and that the heating indeed primarily occurs in the densest gas, i.e., in the post-shock regions. Also note the continuous increase in the overall temperature of the gas between $t=t_\mathrm{cross}$ (left-hand panel) and $t=2\,t_\mathrm{cross}$ (right-hand panel), which leads to the slowly decreasing $\mathcal{M}$ over time.

\section{Summary and conclusion} \label{sec:conclusion}

We performed hydrodynamical simulations of supersonic and subsonic turbulence, employing an ideal equation of state (EOS) with adiabatic indices $\gamma=1.0001$ (nearly isothermal), $\gamma=7/5$ (diatomic molecular gas), and $\gamma=5/3$ (monatomic gas). Section~\ref{sec:results} provided a detailed analysis of the density variance -- Mach number relation, $\sigma_s^2(\mathcal{M})$, which is a key ingredient for theoretical models of the star formation rate and the initial mass function. Unlike previous studies of purely isothermal and polytropic turbulence, we find that an ideal gas EOS leads to a steeper dependence of the density variance $\sigma_s^2$ on the rms sonic Mach number $\mathcal{M}$. We find a new combined approximate relation of the form given by Equation~(\ref{eq:result}) for low Mach numbers, $b\mathcal{M}\lesssim1$, which reduces to the well known isothermal solution for the special case $\gamma\to1$, but also covers cases $\gamma>1$. We argue that the steeper-than-isothermal dependence for $b\mathcal{M}\lesssim1$ is a result of the local heating of the gas in post-shock regions, with the global sonic pre-shock Mach number in the relation being affected later in the evolution. This is because the turbulent driving keeps increasing the internal energy reservoir, leading to an ever increasing global sound speed in adiabatic gases. This is in stark contrast to isothermal and polytropic gases, where the sonic Mach number reaches a statistical steady state rather than continuously decreasing.

We derived a theoretical model, Equation~(\ref{eq:theory}), for the $\sigma_s^2(\mathcal{M})$ relation in Section~\ref{sec:discussion}, which is based on the Rankine-Hugoniot shock jump conditions and provides reasonable fits to all our data. It furthermore predicts a saturation of $\sigma_s^2$ for $b\mathcal{M}>>1$, which is not yet in reach by numerical simulations. Such a saturation is reasonable for $\gamma>1$, given the fact that adiabatic shocks always have a finite jump in density, while isothermal shocks can theoretically have an infinitely large jump in density across the shock. Both Equation~(\ref{eq:result}) for $b\mathcal{M}\lesssim1$ and Equation~(\ref{eq:theory}) for $b\mathcal{M}>1$ naturally simplify to the standard isothermal relation, Equation~(\ref{relation1}) for $\gamma=1$.

We conclude that changes in the adiabatic exponent $\gamma$ can introduce important modifications in the density variance -- Mach number relation and we provide an approximation of that behaviour in Equation~(\ref{eq:result}). However, we emphasize that the real ISM is a mixture of atomic and molecular phases and that the effective EOS is determined by a complex balance of heating and cooling processes, which in turn depend on the chemical evolution and exposure to interstellar and local radiation fields (e.g., from massive stars). Thus, our systematic study sheds some light on the dependence of turbulent density fluctuations on the thermodynamics and composition of interstellar gas and more detailed studies including realistic heating and cooling are required to make further progress.

We hope that this work provides a more general understanding of the density variance -- Mach number relation in the ISM. This is especially true for the warm, atomic part of the ISM, where the gas is clearly not isothermal and may be approximated with an adiabatic equation of state with $\gamma > 1$, which was not covered by previous density variance -- Mach number relations in the literature. Our new relations in this paper attempt to cover this regime and do seem to approximately do so, as tested with the set of simulations presented here. We hope that the new relations will provide useful generalisations of the previous (purely isothermal) relations, which are key ingredients for theoretical models of star formation \citep[see e.g., the review by][]{PadoanEtAl2014}.

\section*{Acknowledgments}
We thank the referee, Sam Falle, for his timely and constructive report, which improved the paper significantly.
C.F.~acknowledges funding provided by the Australian Research Council's Discovery Projects (grants~DP130102078 and~DP150104329).
We gratefully acknowledge the J\"ulich Supercomputing Centre (grant hhd20), the Leibniz Rechenzentrum and the Gauss Centre for Supercomputing (grant pr32lo), the Partnership for Advanced Computing in Europe (PRACE grant pr89mu), and the National Computational Infrastructure (grants~n72~and~ek9), supported by the Australian Government.
This work was further supported by resources provided by the Pawsey Supercomputing Centre with funding from the Australian Government and the Government of Western Australia.

\end{document}